# A Comparative Study on Approaches to Acoustic Scene Classification using CNNs

Ishrat Jahan Ananya[1], Sarah Suad[2], Shadab Hafiz Choudhury[3]*, Mohammad Ashrafuzzaman Khan[4]

[1] North South University, Dhaka, Bangladesh
`ishrat.jahan16@northsouth.edu`
[2] North South University, Dhaka, Bangladesh
`sarah.suad@northsouth.edu`
[3] North South University, Dhaka, Bangladesh
`shadab.choudhury@northsouth.edu`
[4] North South University, Dhaka, Bangladesh
`mohammad.khan02@northsouth.edu`

**Abstract.** Acoustic scene classification is a process of characterizing and classifying the environments from sound recordings. The first step is to generate features (representations) from the recorded sound and then classify the background environments. However, different kinds of representations have dramatic effects on the accuracy of the classification. In this paper, we explored the three such representations on classification accuracy using neural networks. We investigated the spectrograms, MFCCs, and embeddings representations using different CNN networks and autoencoders. Our dataset consists of sounds from three settings of indoors and outdoors environments - thus the dataset contains sound from six different kinds of environments. We found that the spectrogram representation has the highest classification accuracy while MFCC has the lowest classification accuracy. We reported our findings, insights as well as some guidelines to achieve better accuracy for environment classification using sounds.

## 1 Introduction

Acoustic Scene Classification (ASC) is the process of understanding and classifying scenes and environments from ambient audio. It has plenty of use cases in autonomous systems such as monitoring, self-driving vehicles and robotics; in helping those with visual and hearing disabilities understand their surroundings better; and in analysing multimedia recordings. However, there has not been much work done on designing a generic and scalable methodology for solving audio classification problems. The lack of a generic approach is a major issue, as it has limited the development and widespread application of ASC.

In ASC problems, we generally start with unstructured data in the form of audio files. Audio can be represented as 2-D matrices of amplitude, energy, or another property of sound against time. As we have a large volume of unstructured data, Deep Learning approaches such as Convolutional Neural Networks (CNNs) would be very effective at extracting features. While they are commonly used for images, they can also be applied to other forms of data that are in 2-D matrix form. To carry out ASC using CNNs, we must first extract acoustic features such as volume, pitch and sequence of sounds. Raw audio files do not expose these features easily. To get features we can input into a CNN, we must convert an audio file into a different form that represents features better. Three popular approaches are Spectrograms, MFCCs and Embeddings. These approaches are detailed in the Methodology section. Previous research has used these approaches to solve specific problems in the field of ASC. They were able to optimize a particular approach to a very high degree. However, they did not create any kind of general solution that is applicable on a wider scale.

We hypothesized that a more generic approach or technique will improve scalability and reproducibility. Our goal was to figure out what kind of preprocessing and feature representation gives us the best result. This would be the first step in developing a general approach to CNN-based acoustic scene classification. In order to compare the different feature representation approaches, a broad problem was selected: classification of indoor and outdoor scenes. We used an audio dataset with six different types of scenes, converted them into each of the three feature representations mentioned above and compared their performances using several popular CNN models.

The experimental results show that Spectrograms offer the best results, reaching up to 90% accuracy on this problem. The MFCCs were less effective, as the type of features they represent are not so distinguishable when applied to ASC. These two approaches are very commonly used in audio classification. Embeddings are a less common approach, but they proved to be a very lightweight and efficient solution. At around 80% accuracy, the results are acceptable but could be improved further. They could be used in limited devices such as mobile phones where speed and efficiency is more important.

The paper is organized in the following way: first we discuss previous research done in the field of classifying audio using artificial intelligence as well as ASC in general. Then we cover the experimental process, starting from the preparation of the data and details on the CNN models used. Finally, we evaluate and discuss the results of the experiments, and put forward some suggestions for future work.

## 2 Related Work

Analysis and classification of auditory signals with artificial intelligence have a long history. Initially, research work was focused on simply detecting and distinguishing acoustic events [1] such as distinct noises like claps and speech, or different individuals speaking [2]. These early examples of the use of neural networks in the classification of audio developed from the intersection of signal processing and artificial intelligence.

More mature artificial intelligence techniques such as sophisticated convolutional neural networks have enabled further exploration of Acoustic Scene Classification through different approaches. The DCASE Challenges, initially started in 2013, offer datasets and a platform for the exploration of Acoustic Scene Classification [3]. DCASE 2013 highlighted the use of large datasets for acoustic scene classification in various scenes such as a bus, office, market, et cetera.

Early research yielded good outcomes with machine learning models. Good results were achieved in the DCASE 2013 challenge using algorithms such as support vector machines and decision trees [4]. However, as the sophistication and size of datasets increased, neural networks became an effective choice. In this specific dataset, Valenti et al's approach using a custom CNN model resulted in higher accuracy compared to earlier work - up to 9.7% depending on the technique it is compared to [5].

The majority of recent work on acoustic scene classification has followed up on the CNN approach. Hussein et al. developed a more in-depth technique using a deep neural network with only 3 hidden layers that achieved up to 90% accuracy on the DCASE 2016 challenge [6].

In the DCASE 2020 challenge, several attempts were able to reach 96% test accuracy by implementing modern deep convolutional neural networks such as ResNets. While both these and the previous approach were highly accurate, they also made use of specific preprocessing techniques and model designs that could be difficult to implement on a larger scale.

Finally, an excellent overview of the development and use of deep learning in acoustic scene classification between 2013 and 2020 is given in a review paper by Abeßer [7].

## 3 Methodology

The first step is to convert the audio file, an uncompressed .wav file, into a numerical form. This returns a one-dimensional array of length equal to the sampling rate in hertz – in this case 44,100. This array is completely impractical to use in any kind of deep learning application, so we must use a different representation. Initial audio preprocessing was carried out on this data, and then it was converted to the three different feature representations.

The CNN models utilized in this experiment were a simple autoencoder, AlexNet, ResNet-18 and ResNet-50. ResNets were used because they are an extremely robust and powerful architecture for classification problems. Due to the number of layers, it would be able to extract features from inputs like Spectrograms that don't have many obvious features. While they are mainly used for images, they could also be used for other types of inputs with acceptable performance. The number of deep layers in ResNet makes it extremely suitable for generalizing, which is one of our goals. Alterations to the structure of the models were kept to a minimum in order to ensure a fair comparison. Any changes made were in order to ensure that the input could be appropriately fed into the model.

### 3.1 Data Organization and Collection

The dataset was collected from DCASE 2020 challenge [8]. It had three classes of indoor, outdoor and transport. These classes were subdivided into nine more subclasses. The raw dataset contained 10 second audio clips in 24-bit .wav format taken from ten different cities in the world.

For this paper, we chose to reduce it to two classes: indoor and outdoor. This left us with six subclasses, three from each. For Indoor Scenes, the subclasses were 'Metro Station', 'Shopping Mall' and 'Airport'. For Outdoor Scenes, the subclasses were 'Park', 'Pedestrian Street' and 'Public Square'.

There were 8640 data samples in total, adding up to 24 hours of audio. Each subclass had 1440 data points. We ensured that the number of audio samples for each class was equal so that the model was not biased towards any particular category due to unbalanced data. The original audio files were binaural at 44.1 kHz. This is a relatively small dataset, so we split each of the 10 second audio files into two 5 second files to double the size of the dataset. Table 1. Gives the size of the dataset at different stages of collection and preprocessing.

**Table 1.** Data Organization

| Type of Data | No. of Data Points | Hours of Audio |
| --- | --- | --- |
| Files in Dataset (10-sec clips) | 8,640 | 24 |
| Files per subclass | 1,440 | 4 |
| Indoor Scenes (5-sec clips) | 8,640 | 12 |
| Outdoor Scenes (5-sec clips) | 8,640 | 12 |
| Indoor Scenes with Augmentation | 17,280 | 24 |
| Outdoor Scenes with Augmentation | 17,280 | 24 |
| Test Data per Class | 3,456 | 4.8 |
| Train Data per Class | 13,824 | 19.2 |
| Total Data (with 2 classes) | 34,560 | 48 |

## 3.2 Data Augmentation

Before converting the data into a feature representation, some data augmentation was carried out. A small amount of random noise was added and audio tracks were randomly shifted forwards and backwards.

Many other typical augmentations, such as pitch-shifting and extending silences, were determined to be detrimental to this dataset. They are typically used for speech-based datasets. Here, the audio is in the form of a continuous stream of noise rather than speech at different tones and with small silences in between.

This gave us a rich dataset of 34,560 five-second audio clips, half of which was augmented.

## 3.3 Feature Representations

The final part of data preprocessing required taking audio samples as input and extracting features from the audio signals. By doing so, we aim to find components of the audio signals that will help us differentiate it from other categories of signals. We implemented three methods of generating feature representations: producing log-mel spectrograms [9], Mel Frequency Cepstral Coefficients (MFCC) [10] and audio embeddings.

**MFCCs and Spectrograms.**
To generate MFCCs, first, the audio signal is sliced into 20ms wide frames. We assume that there is no change in the signal within each 20ms frame. We applied a short-term Fourier Transform on each frame to calculate the power spectrum. This gives us the distribution of power into frequency components that make up the signal. Next, we applied the Mel filter bank to each power spectra and summed up the energy in each filter. This step actually estimates how the human ear perceives sounds at different frequencies and different volumes. The last step in the process is to take the discrete cosine transform of the logarithm of each filter bank. All these calculations were carried out using the python library Librosa [11].

Log-mel spectrograms were produced in a similar way. The audio dataset underwent a short-time Fourier transform to get spectrograms based on the Frequency and Amplitude of the signal rather than Power. The spectrograms were then scaled to the Mel scale and saved in .png format.

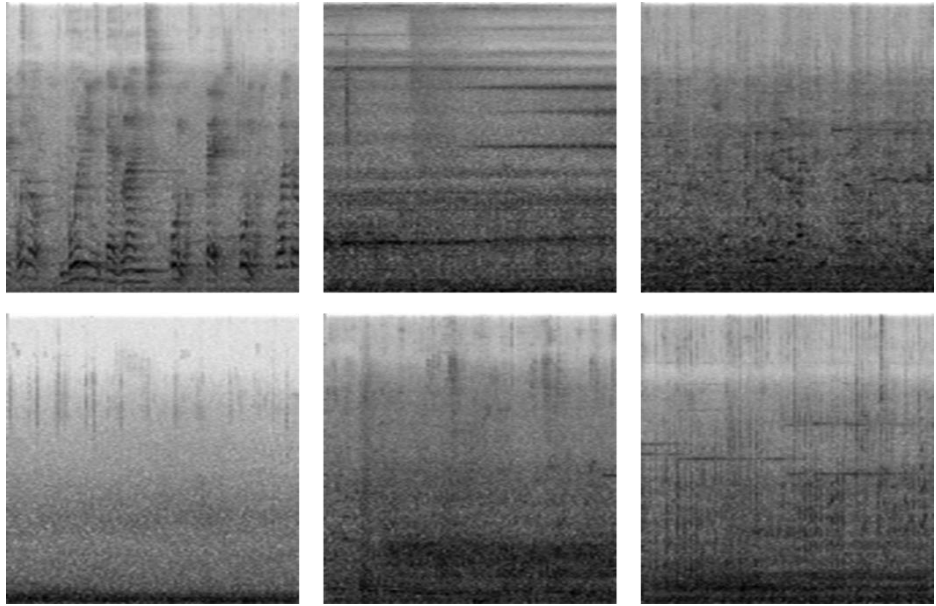

**Fig. 1.** Spectrograms of each class. The top row shows Indoor (Airport, Metro Station and Shopping Mall from left to right). The bottom row shows Outdoor (Park, Public Square and Pedestrian Street from left to right).

**Audio Embeddings**

Humans categorize or recognize things by comparing its details to previous knowledge. Many image classification algorithms use the same approach. These algorithms are trained on datasets where the input is in the form of images that have objects labelled in them. Using audio embeddings utilizes a similar approach for audio classification.

Embeddings are used to map items from a high dimensional vector space to a low dimensional vector space. In dense data such as audio, the embeddings determine similarity metrics to other sounds. Essentially, it splits the audio clip into smaller intervals. For each interval, it gives a similarity value to all the classes the original embedding model was trained on. So, while MFCCs and Spectrograms use features extracted directly from the audio signals, Embeddings simply use learned features generated by another machine learning model that has been trained to label audio files.

The embeddings we used were generated using an Audio Embedding Generator [12, 13]. The generator accepts a 16-bit PCM .wav file as input, embeds the feature labels and outputs the result as arrays of 1 second embeddings. The model was trained on Audioset, which includes 632 classes. For each second, the embeddings list the 128 classes that have the highest similarity to the sound.

The following figure shows an example of each type of preprocessed data, in a visual format. Note, that while the MFCCs and Embeddings are depicted visually here, they were input into the CNN as two-dimensional matrices.

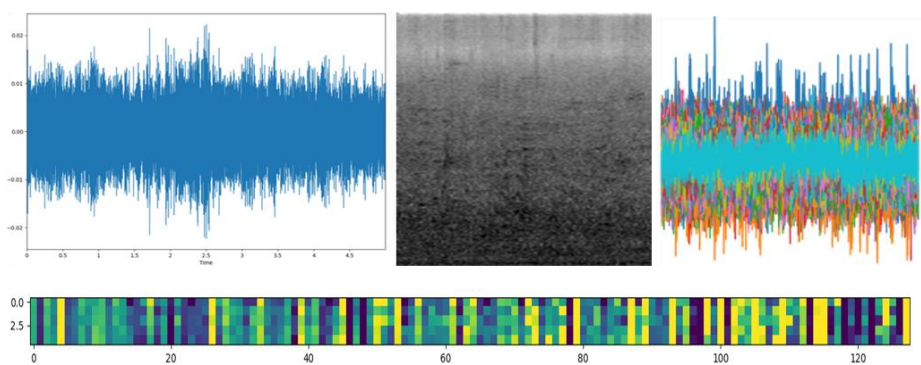

**Fig. 2.** Four representations of an audio file recorded at an Airport in Lisbon: a waveform plot (top left), a Spectrogram (top middle), MFCCs plotted as a graph (top right) and Audio Embeddings plotted as a graph (bottom).

### 3.4 Development of CNNs

Inputting the Spectrograms into the CNNs for training was straightforward, as the spectrograms were saved as 224x224-sized images. The MFCCs and Embeddings were in the form of two-dimensional matrices, and so could not be input directly. They had to be reshaped before being fed into the CNNs. When necessary, the initial layer was altered to fit the dataset.

We tested four CNNs for all three feature representations: ResNet-18, ResNet-50, AlexNet and finally an Autoencoder. For each CNN, a limited degree of hyperparameter optimization was carried out.

The ResNet family of neural network architectures is ideal for image classification tasks. The ResNet architecture uses stacked layers of residual learning blocks using shortcuts between layers to minimize the effect of the vanishing gradient problem [15]. We used the SGD optimizer with Cross-Entropy Loss and a learning rate of 0.001. Despite being an older architecture, AlexNet [16] was also tested to see how the number of parameters affects the results. The results of both models are discussed in the Evaluation section.

We initially tested the autoencoder model for the Embeddings only. The structure of the embeddings is such that labels for similar sounds would be clustered together. Using a simple Autoencoder with linear layers enhanced this and exposed the largest clusters. For the sake of comparison, other autoencoder models were also tested on the other two feature representations.

## 4 Results and Evaluation

The following table gives a summary of the results of our experimentation, and a discussion of the results follows.

Table 2. Experimental Results for Spectrograms

| CNN Architecture | Test Accuracy | Training Accuracy |
|---|---|---|
| Autoencoder | 76.2% | 79.3% |
| AlexNet | 86.3% | 90.7% |
| ResNet-18 | 89.7% | 91.9% |
| ResNet-50 | 90.4% | 93.6% |
| ShuffleNet | 93.1% | 95.2% |

It's clear from the table that the Spectrograms offer the best results, up to 93%. Considering that these CNNs are normally used for image classification, these results are as expected. Most of the models give somewhat acceptable results with the other feature representations, getting around 70-80% accuracy.

|  | predicted | | |
|---|---|---|---|
| n = 3456 | Indoor | Outdoor | |
| actual Indoor | 1575 | 153 | 1728 |
| actual Outdoor | 88 | 1640 | 1728 |
| | 1728 | 1728 | |

**Fig. 3.** Confusion Matrix for the best result – 93.1% with ShuffleNet

Even a simple fully-connected autoencoder of 4096-2048-1024-512 parameters gives us 76.3% accuracy, implying that the spectrograms are actually shallow and do not have a lot of features to extract. We used the encoder layer from the aforementioned autoencoder to improve feature extraction before feeding the parameters into the ResNets to gain a slight improvement. After getting an extremely high value from the ResNets but a lower value from AlexNet, we decided to try another state-of-the-art model with fewer parameters. We chose Shufflenet, which is a computationally efficient CNN architecture, particularly designed for mobile devices with limited processing power [17]. While these results do not break any of the benchmarks set in previous DCASE challenges, they are all fairly generic approaches that require minimal customization to the dataset. This ensures that the results are applicable across different acoustic scene datasets rather than being optimized for this particular problem.

**Table 3.** Experimental Results for MFCCs

| CNN Architecture | Test Accuracy | Training Accuracy |
|---|---|---|
| Autoencoder | 49.9% | 50.0% |
| AlexNet | 69.8% | 89.1% |
| ResNet-18 | 71.3% | 86.6% |
| ResNet-50 | 72.1% | 88.0% |

Due to the features of the MFCCs, using an Autoencoder was completely ineffective, as shown by the 50% accuracy on a binary classifier. We note that ResNet models are actually too heavy for the MFCC, generally overfitting within a few epochs. The text accuracies recorded are before it fully overfits. Using the non-augmented half of the data set saw a slight improvement in accuracy, at the cost of even more overfitting. At this stage, regularization techniques were ineffective. Therefore, for a dataset of this size, augmentation was necessary.

MFCCs are generally used for speech classification. It can easily distinguish between high and low volumes and pitches. However, the audio in this dataset is primarily background noise at a similar energy level throughout. Therefore, it is harder to extract features using MFCCs compared to other approaches and this feature representation has the lowest accuracy of all.

**Table 4.** Experimental Results for Embeddings

| CNN Model | Test Accuracy | Training Accuracy |
|---|---|---|
| Autoencoder | 80.8% | 82.2% |
| AlexNet | 77.9% | 96.8% (overfit) |
| ResNet-18 | 77.6% | 99.7% (overfit) |
| ResNet-50 | 77.1% | 99.6% (overfit) |

The audio embeddings performed surprisingly well considering the nature of the dataset. Reducing the audio dataset into a series of labels allowed the autoencoder to learn features very easily even if there was much less data compared to the other approaches. The original dataset used to develop the embedding generator was focused on speech, music and the sounds made by individual objects. Despite being a somewhat unsuitable dataset, it gave good results. With a more closely related embedding generator, it could give results comparable to Spectrograms at a fraction of the computation power. However, no such generator for urban scenes is currently available, and developing one from scratch would be out of the scope of this paper. Finally, as seen in the table, heavier models like the ResNets and AlexNet led to overfitting when used with embeddings.

At this point, it should be noted that all the models that were run on individual subclasses faced significant issues. Even with plenty of augmentation, classifying on six classes rather than two led to extensive overfitting and an accuracy of 70% at most. It is clear that the dataset is too small to achieve good accuracy unless the classes are combined. Generally, the models trained on the Embeddings were able to distinguish Metro Station, Shopping Mall, Park and Public Square with a high degree of accuracy – close to 80%, while Airport and Pedestrian Street displayed lower confidence. Models trained on the Spectrograms showed relatively similar confidence per subclass.

With typical image classification problems, different objects or segments are highly visible, as they have clear contours and different colours. The Spectrograms are not so distinguishable, as seen in Fig. 1. So, a huge amount of data is needed to train the models to be able to distinguish between different classes. The MFCCs and Embeddings do not suffer from this issue, but are not suitable for typical deep networks and may require custom CNNs to improve further.

Additionally, it is clear from both our work and other research that, unlike image classification, audio classification does not always benefit from deep networks with lots of parameters. The number of features that can be extracted from the Spectrograms or other formats is extremely limited. AlexNet has 61 million parameters, while ResNet-18 has 11 million and ResNet-50 has 23 million. All three networks have very similar accuracies and are in fact prone to overfitting. By contrast, the smallest network used here, the fully-connected autoencoder with audio embeddings, only has 560k parameters. Since datasets for ASC are fairly limited, adding more data to solve this problem is not always feasible. Therefore, higher input resolution [18] or additional preprocessing may be necessary to achieve better accuracy.

## 5   Conclusion

As the evaluation section shows, the task of Acoustic Scene Classification faces major hurdles when it comes to larger models. Extensive data preprocessing and augmentations are necessary to achieve high accuracies on even very limited problems. Out of the three different approaches tested, Spectrograms offered the best result for acoustic scene classification of interior and exterior urban scenes. It achieved over 90% accuracy, but required a lot of data to reach this accuracy.

We suggest not using MFCCs for ASC, as they require extensively customized CNNs to get a good result.

For a lightweight approach, audio embeddings are suitable. Even though the embedding generator model was not completely suitable for this domain, it was able to achieve almost 81% accuracy. Higher accuracy could easily be reached if an embedding generator model is developed that is focused on urban audio. We suggest focusing on this approach for further development of a generic approach, since it is an efficient process suitable for use in low-power mobile devices.

Future work to follow up on this paper would involve two other approaches. The first would be to increase the resolution and accuracy of our comparison of models by using a wider selection of models to see which approach to audio classification works best at different sizes of models. Secondly, there are several datasets available for ASC outside DCASE. Combining multiple datasets may enable more general conclusions to be drawn.

These avenues of future development mentioned will place the groundwork for CNNs focused on audio classification and help acoustic scene classification to be used in wider contexts.